\documentclass[conference]{IEEEtran}
\IEEEoverridecommandlockouts

\usepackage{amsmath,amsopn,amssymb,bm}
\usepackage{amsfonts}
\usepackage{amsthm} 
\usepackage{graphicx,xspace,color}
\usepackage{epsfig}
\usepackage{longtable,multirow}
\usepackage{array}
\usepackage{stfloats}
\usepackage{cuted}
\usepackage{cite}
\usepackage[nolist]{acronym}
\usepackage{soul}
\usepackage{algorithm}
\usepackage[noend]{algpseudocode}
\usepackage{algcompatible}
\usepackage{enumerate}
\usepackage{lettrine}
\usepackage{mathtools}
\usepackage{subcaption}
\usepackage{comment}
\usepackage{stfloats}
\usepackage{gensymb}
\usepackage{hyperref}

\usepackage[justification=centering]{caption}

\usepackage{booktabs}
\def\tablename{Table}
\graphicspath{ {./figures/}}
\newtheorem*{remark*}{Remark}
\newtheorem{theorem}{Theorem}

\makeatletter

\def\ps@IEEEtitlepagestyle{
	\def\@oddfoot{\mycopyrightnotice}
	\def\@evenfoot{}
}
\def\mycopyrightnotice{
	{\footnotesize \copyright~2020 IEEE. Personal use is permitted, but republication/redistribution requires IEEE permission. See \href{https://www.ieee.org/publications/rights/index.html}{IEEE Intellectual Property Rights} for more information.\hfill} 
	\gdef\mycopyrightnotice{}
}

\begin{document}
	\begin{acronym}
		\acro{MIMO}{multiple-input multiple-output}
		\acro{BS}{base station}
		\acro{LoS}{line-of-sight}
		\acro{NLoS}{non-line-of-sight}
		\acro{MMSE}{minimum mean square error}
		\acro{SE}{spectral efficiency}
		\acro{MR}{maximum-ratio}
		\acro{MU-MIMO}{multiuser MIMO}
		\acro{UatF}{use-and-then-forget}
		\acro{UL}{uplink}
		\acro{SNR}{signal-to-noise ratio}
		\acro{TDD}{time-division-duplex}
	\end{acronym}

	\title{Massive MIMO with Multi-Antenna Users under Jointly Correlated Ricean Fading}
	
	\author{\IEEEauthorblockN{Konstantinos Dovelos\IEEEauthorrefmark{1}, Michail Matthaiou\IEEEauthorrefmark{2}, Hien Quoc Ngo\IEEEauthorrefmark{2}, and Boris Bellalta\IEEEauthorrefmark{1}}
		\IEEEauthorblockA{$^*$Department of Information and Communication Technologies, Universitat Pompeu Fabra (UPF), Barcelona, Spain}
		\IEEEauthorblockA{$^\dagger$Institute of Electronics, Communications and Information Technology (ECIT), Queen's University Belfast, Belfast, U.K.}
		Email: \{konstantinos.dovelos, boris.bellalta\}@upf.edu, \{m.matthaiou, hien.ngo\}@qub.ac.uk}
	\maketitle

\begin{abstract}
We study the uplink performance of massive \ac{MIMO} when users are equipped with multiple antennas. To this end, we consider a generalized channel model that accounts for line-of-sight propagation and spatially correlated multipath fading. Most importantly, we employ the Weichselberger correlation model, which has been shown to alleviate the deficiencies of the popular Kronecker model. The main contribution of this paper is a rigorous closed-form expression for the uplink spectral efficiency using maximum-ratio combining and minimum mean square error channel estimation. Our result is a non-trivial generalization of previous results on massive MIMO with spatially correlated channels, thereby enabling us to have suitable designs for future massive MIMO systems. Numerical simulations corroborate our analysis and provide useful insights on how different propagation conditions affect system performance. 
	\end{abstract}

	\section{Introduction}
	Massive \acf{MIMO} is now a mature technology, and has become an integral component of 5G communication systems. Deploying a massive number of \ac{BS} antennas to serve multiple users over the same time-frequency resources can yield huge \ac{SE} gains, whilst simple linear processing techniques are nearly optimal \cite{marz1}, \cite{marz2}. 
	
	There is a large body of literature investigating the performance of \ac{MU-MIMO} under spatially correlated fading channels. Most of them invoke the Kronecker (or separately-correlated) model \cite{kro_model}, which enforces the spatial correlation properties at the transmitter and the receiver to be separable. Kronecker-type models are analytically tractable, however they have been shown to inadequately represent a variety of practical channels, such as indoor \ac{MIMO} channels~\cite{kron_deficiencies}. A more realistic model is the so-called Weichselberger model~\cite{wei_model}, which alleviates the deficiencies of the Kronecker model by jointly accounting for the correlation at both link ends and their mutual dependence. Modeling accurately spatial correlation in \ac{MU-MIMO} is therefore essential, since simplistic channel models may wrongly estimate the actual system performance and, consequently, cannot be leveraged for practical transceiver design. 
	
	To this end, \cite{massive_distri_centr} considered both centralized and distributed massive \ac{MIMO} using the Weichselberger model, but did not account for \ac{LoS} propagation. More recently, \cite{WModel_SingleAntenna}~investigated the performance of massive \ac{MIMO} with single-antenna users in terms of favorable propagation and channel hardening under \ac{LoS} and spatial correlation based on the Weichselberger model. Furthermore, a stream of recent papers (see \cite{single_antenna_ricean1}, \cite{single_antenna_ricean2}, and references therein) analyzed the performance of multi-cell massive \ac{MIMO} with single-antenna users under correlated Ricean fading and pilot contamination. 
	
	Even though most of the user devices today are equipped with multiple antennas, existing studies on massive \ac{MIMO} mainly focus on single-antenna users. In the related literature, we distinguish the work in \cite{massive_mimo_multiantenna_users}, which studied massive \ac{MIMO} with multi-antenna users, though using the Kronecker model, neglecting \ac{LoS} propagation, and assuming precoding at the users. To the best of our knowledge, massive \ac{MIMO} has not been studied yet under a generalized setup where users are equipped with multiple antennas and spatial correlation is described jointly at both link ends. 
	
	This paper aims to fill this gap in the literature by analyzing the \ac{UL} performance of massive \ac{MIMO} with multi-antenna users under jointly correlated Ricean fading. Most importantly, in our channel model, spatial correlation is described at the transmitter and the receiver jointly by the Weichselberger model, which includes the popular Kronecker and virtual channel representation \cite{vcr_model} models as special cases. Our analysis then focuses on the ergodic sum \ac{SE}. By employing the  \ac{UatF} method~\cite{mMIMO_book}, we compute a lower bound on the system capacity. Next, we derive a closed-form expression for this bound, which represents an achievable \ac{SE} using \ac{MMSE} channel estimation, \ac{MR} combining, and per-stream decoding. We finally conduct Monte Carlo simulations to verify our analysis and study the system performance under various propagation conditions. More particularly, we investigate if the sum \ac{SE} grows always without bound in jointly correlated Ricean fading as the number of \ac{BS} antennas increases, as well as whether it is better to serve many single-antenna users or few multi-antenna users. Note that rigorous answers to these fundamental questions are, in general, not available in the literature for such generalized channel models.
	
	\textit{Notation}: $\mathbf{A}$ is a matrix; $\mathbf{a}$ is a vector; $[\mathbf{A}]_{i,j}$ is the $(i,j)$-th entry of $\mathbf{A}$; $(\cdot)^*$, $(\cdot)^T$, and $(\cdot)^H$ denote conjugate, transpose, and conjugate transpose, respectively; $\text{vec}(\mathbf{A})$ is the column vector formed by the stack of the columns of $\mathbf{A}$; $\otimes$ and $\odot$ denote the Kronecker and the element-wise products, respectively; $\text{tr}(\cdot)$ is the trace; $\|\cdot\|_{\textsc{F}}$ is the Frobenious norm; $\mathbf{I}_N$ is the $N\times N$ identity matrix; $\mathbf{1}_{M\times N}$ is the $M \times N$ matrix with unit entries; $\mathcal{CN}(\bm{\mu},\bm{\Sigma})$ is a complex Gaussian vector with mean $\bm{\mu}$ and covariance matrix $\bm{\Sigma}$; $\mathcal{U}(a,b)$ denotes the uniform distribution over the interval $(a,b)$; $\mathbb{E}[\cdot]$ denotes expectation; and $\text{Re}\{\cdot\}$ is the real part of a complex variable.
	
	\vspace{-4pt}
	\section{System Model}\label{sec:system_model}
		Consider the \ac{UL} of a massive MIMO system operating in \ac{TDD} mode, where the BS serves $K$ users in the same time-frequency resource. The BS is equipped with $M$ antennas while users have $N$ antennas each.
		
	\subsection{The Weichselberger Ricean Fading Channel}
	The channel from the $k$-th user to the \ac{BS} is denoted by $\mathbf{H}_k=[\mathbf{h}_{k1},\dots,\mathbf{h}_{kN}]\in\mathbb{C}^{M\times N}$. We next consider jointly correlated Ricean fading, where $\text{vec}(\mathbf{H}_k)\sim\mathcal{CN}(\text{vec}(\bar{\mathbf{H}}_k), \mathbf{R}_k)$. The matrix $\bar{\mathbf{H}}_k=[\bar{\mathbf{h}}_{k1},\dots,\bar{\mathbf{h}}_{kN}]$ corresponds to the deterministic \ac{LoS} component, whereas $\mathbf{R}_k$ follows the Weichselberger model~\cite{wei_model}. The channel of user $k$ is hence expressed as
	\begin{align}\label{eq:channel_model}
	\mathbf{H}_k = \bar{\mathbf{H}}_k  + \underbrace{\mathbf{U}_{k,r}(\tilde{\bm{\Omega}}_k\odot\mathbf{H}_{\text{iid}})\mathbf{U}_{k,t}^T}_{\tilde{\mathbf{H}}_k}
	\end{align}
	where $\tilde{\mathbf{H}}_k$ is the stochastic \ac{NLoS} component, $\tilde{\bm{\Omega}}_k$ is an $M\times N$ deterministic matrix with real-valued nonnegative elements, and $\mathbf{H}_{\text{iid}}$ is a matrix whose entries are independent and identically distributed (i.i.d.) $\mathcal{CN}(0,1)$. The unitary matrices $\mathbf{U}_{k,r}\in\mathbb{C}^{M\times M}$ and $\mathbf{U}_{k,t}\in\mathbb{C}^{N\times N}$ are the eigenbases of the one-sided correlation matrices $\mathbf{R}_{k,r}\triangleq\mathbb{E}[\tilde{\mathbf{H}}_k\tilde{\mathbf{H}}_k^H]$ and $\mathbf{R}_{k,t} \triangleq \mathbb{E}[\tilde{\mathbf{H}}_k^T\tilde{\mathbf{H}}_k^*]$, respectively. Let $\bm{\lambda}_{k,r}$ and $\bm{\lambda}_{k,t}$ denote the vectors of the eigenvalues of the one-sided correlation matrices. Due to the joint correlation feature of the channel, the eigenvalues are coupled through the constraints $[\bm{\lambda}_{k,r}]_{m} = \sum_{n=1}^{N}\left[\bm{\Omega}_{k}\right]_{m,n}$ and $[\bm{\lambda}_{k,t}]_{n}  = \sum_{m=1}^{M}\left[\bm{\Omega}_{k}\right]_{m,n}$, where $\bm{\Omega}_k \triangleq \tilde{\bm{\Omega}}_k\odot\tilde{\bm{\Omega}}_k$ is the so-called \textit{eigenmode coupling} matrix. The real-valued and nonnegative element $[\bm{\Omega}_k]_{m,n}$ specifies the average power coupling between the $n$-th transmit eigenmode and the $m$-th receive eigenmode of link $k$. Given the eigenbases and the coupling matrix, one can compute the full correlation matrix $\mathbf{R}_k$ as
	\begin{align}\label{eq:correlation_matrix}
	\mathbf{R}_k &\triangleq \mathbb{E}\left[\text{vec}(\tilde{\mathbf{H}}_k)\text{vec}(\tilde{\mathbf{H}}_k)^H\right] \nonumber\\
	&=(\mathbf{U}_{k,t}\otimes\mathbf{U}_{k,r})\text{diag}(\text{vec}(\bm{\Omega}_k))(\mathbf{U}_{k,t}\otimes\mathbf{U}_{k,r})^H
	\end{align}
	where $\text{diag}(\cdot)$ is the diagonal matrix formed by the elements of the input vector. Referring to \eqref{eq:channel_model}, the Ricean factor $\kappa_k$ and the large-scale fading coefficient $\beta_k$ of user $k$ are incorporated into the model through the constraints $\|\bar{\mathbf{H}}_k\|^2_F= MN\beta_k\frac{\kappa_k}{\kappa_k+1}$ and $\text{tr}(\mathbf{R}_k) = MN\beta_k\frac{1}{\kappa_k+1}$ \cite{SR_WModel}. Hereafter, we neglect the effect of large-scale fading as our main focus is on modeling the small-scale fading variations; hence, we set $\beta_k=1, \forall k=1\dots,K$.
	
	\begin{remark*}
	Setting $\bm{\Omega}_{k} = \frac{\kappa_k+1}{MN}\bm{\lambda}_{k,r}\bm{\lambda}_{k,t}^T$ yields the Kronecker model~\cite{wei_model}.
	\end{remark*}

	\subsection{Channel Estimation}
	We assume a block fading model where channel responses remain constant over a coherence block of $\tau_c$ symbols. The \ac{BS} exploits \ac{TDD} reciprocity and estimates the channels through uplink pilots sent by the users during a training phase. We point out that downlink transmissions are neglected. Let $\mathbf{P}_k\in\mathbb{C}^{N\times \tau_p}$ denote the pilot matrix of user $k$ with $\mathbf{P}_k\mathbf{P}_{k'}^H = \delta_{kk'}\mathbf{I}_N$, where $ \delta_{kk'}$ denotes the Kronecker delta function. The training phase spans $\tau_p\geq KN$ symbols \cite{training_mimo}. Then, each user~$k$ transmits the pilot signal $ \sqrt{\tau_p\rho_p}\mathbf{P}_k$ where $\rho_p$ is the \ac{SNR} of each pilot symbol. After $\tau_p$ channel uses, the \ac{BS} receives 
	\begin{equation}
	\mathbf{Y} = \sqrt{\tau_p\rho_p}\sum_{k=1}^K\mathbf{H}_{k}\mathbf{P}_k + \mathbf{N} 
	\end{equation}
	where $\mathbf{N}\in\mathbb{C}^{M\times\tau_p}$ is the normalized noise matrix with i.i.d. $\mathcal{CN}(0,1)$ entries. The BS is assumed to have perfect knowledge of the channel statistics of all users, namely of the mean $\text{vec}(\bar{\mathbf{H}}_k)$ and the covarianc matrix $\mathbf{R}_k$.\footnote{This is a reasonable assumption as the channel statistics change over a much longer timescale spanning tens of coherence blocks.} Hence, the BS obtains the \ac{MMSE} estimate of $\mathbf{H}_k$ as $\hat{\mathbf{H}}_k=[\hat{\mathbf{h}}_{k1},\dots,\hat{\mathbf{h}}_{kN}]$, where
	\begin{align*}
	\text{vec}(\hat{\mathbf{H}}_{k}) = \text{vec}(\bar{\mathbf{H}}_{k}) + \sqrt{\tau_p\rho_p}\mathbf{R}_{k}(\tau_p\rho_p\mathbf{R}_{k} + \mathbf{I}_{MN})^{-1}\text{vec}(\tilde{\mathbf{Y}}_k)
	\end{align*}
	and $\tilde{\mathbf{Y}}_k = \mathbf{Y}\mathbf{P}_k^H - \sqrt{\tau_p\rho_p}\bar{\mathbf{H}}_{k}$~\cite{paper_ce}. Let $\mathbf{E}_k\triangleq\mathbf{H}_{k} - \hat{\mathbf{H}}_{k}$ be the estimation error matrix of user $k$. The covariance matrix of the channel estimation error is given by
	\begin{align}
	\mathbf{C}_k &\triangleq \mathbb{E}[\text{vec}(\mathbf{E}_k)\text{vec}(\mathbf{E}_k)^H] \nonumber\\
	&= \mathbf{R}_{k} - \tau_p\rho_p\mathbf{R}_{k}(\tau_p\rho_p\mathbf{R}_{k} + \mathbf{I}_{MN})^{-1}\mathbf{R}_{k}.
	\end{align}
	The \ac{MMSE} estimate $\text{vec}(\hat{\mathbf{H}}_{k})$ and the estimation error $\text{vec}(\mathbf{E}_k)$ are independent random vectors distributed as
	\begin{align*}
	\text{vec}(\hat{\mathbf{H}}_{k})&\sim\mathcal{CN}(\bar{\mathbf{H}}_{k}, \mathbf{R}_{k} - \mathbf{C}_k),\\
	\text{vec}(\mathbf{E}_{k})&\sim\mathcal{CN}(\mathbf{0}, \mathbf{C}_k).
	\end{align*}
	To capture the correlation of the channels and the channel estimates of user $k$, we partition $\mathbf{R}_{k}$ into the block form 
	\\
	\begin{equation}\label{eq:correlation_blockform}
	\mathbf{R}_k =  \begin{pmatrix}
	\mathbf{R}^k_{11}  & \mathbf{R}^k_{12}   &\cdots &\mathbf{R}^k_{1N} \\
	\mathbf{R}^k_{21}  &\mathbf{R}^k_{22}    &\cdots &\mathbf{R}^k_{2N}   \\
	\vdots & \vdots  &\ddots & \vdots \\
	\mathbf{R}^k_{N1}  &\mathbf{R}^k_{N2}   &\cdots & \mathbf{R}^k_{NN} 
	\end{pmatrix} 
	\end{equation}
	\\
	and $\mathbf{C}_k$ into the block form 
	\\
	\begin{equation}\label{eq:covariance_error_blockform}
	\mathbf{C}_k =  \begin{pmatrix}
	\mathbf{C}^k_{11}  & \mathbf{C}^k_{12}   &\cdots &\mathbf{C}^k_{1N} \\
	\mathbf{C}^k_{21}  &\mathbf{C}^k_{22}    &\cdots &\mathbf{C}^k_{2N}   \\
	\vdots & \vdots  &\ddots & \vdots \\
	\mathbf{C}^k_{N1}  &\mathbf{C}^k_{N2}   &\cdots & \mathbf{C}^k_{NN} 
	\end{pmatrix} 
	\end{equation}
	\\
	where $\mathbf{R}^k_{ij} \triangleq \mathbb{E}[\mathbf{h}_{ki}\mathbf{h}_{kj}^H] - \bar{\mathbf{h}}_{ki} \bar{\mathbf{h}}_{kj}^H$, and $\mathbf{C}^k_{ij} \triangleq \mathbb{E}[\mathbf{e}_{ki}\mathbf{e}_{kj}^H]$. Finally, we define the cross-covariance matrix of the channel estimates $\hat{\mathbf{h}}_{ki}$ and $\hat{\mathbf{h}}_{kj}$ of user $k$ as $\bm{\Phi}^k_{ij} = \mathbf{R}^k_{ij} - \mathbf{C}^k_{ij}$.

	\subsection{Linear Combining}
	We consider omnidirectional uplink transmissions. More specifically, each user $k$ transmits the signal $\sqrt{\rho_{\text{ul}}}\mathbf{s}_k$, where $\mathbf{s}_k=[s_{k1},\dots,s_{kN}]\sim\mathcal{CN}(\mathbf{0},\mathbf{I}_N)$ is the vector of the data symbols, and $\rho_{\text{ul}}$ denotes the average \ac{SNR} of each data symbol. The received signal at the BS is then written as
	\begin{align}\label{eq:rx_signal}
	\mathbf{y} = \sqrt{\rho_{\text{ul}}}\sum_{k=1}^K\sum_{i=1}^N\mathbf{h}_{ki}s_{ki} + \mathbf{n}
	\end{align}
	where $\mathbf{n}\sim\mathcal{CN}(\mathbf{0},\mathbf{I}_M)$ is the normalized noise vector. The \ac{BS} employs single-stream decoding and treats the $N$ data streams of user $k$ as being transmitted by $N$ independent single-antenna users. Then, the symbol $s_{ki}$ of user $k$ is detected based on the post-processed signal
	\begin{align}\label{eq:estimated_symbol}
	\tilde{y}_{ki} = \mathbf{w}_{ki}^H\mathbf{y} 
	\end{align}
	where $\mathbf{w}_{ki} = \hat{\mathbf{h}}_{ki}$ is the \ac{MR} combiner for stream $i$. 
	
	\section{Spectral Efficiency Analysis}\label{section:se_analysis}
	It is difficult to compute the maximum achievable \ac{SE} when the receiver has imperfect channel knowledge~\cite{imperfect_csi}. Therefore, we resort to a common bounding technique in massive \ac{MIMO} called \ac{UatF} \cite{mMIMO_book}, which yields achievable yet suboptimal rates. Specifically, adding and subtracting the expected value of the effective channel $\mathbf{w}_{ki}^H\mathbf{h}_{ki}$ in \eqref{eq:estimated_symbol} yields
	\begin{align}\label{eq:post_signal}
	&\tilde{y}_{ki} = \sqrt{\rho_{\text{ul}}}\mathbb{E}[\mathbf{w}_{ki}^H\mathbf{h}_{ki}] s_{ki}+ \sqrt{\rho_{\text{ul}}}\sum_{\substack{m=1, m\neq k}}^K\sum_{j=1}^N\mathbf{w}_{ki}^H\mathbf{h}_{mj}s_{mj}\notag\\ 
	&+\sqrt{\rho_{\text{ul}}}\sum_{\substack{j=1, j\neq i}}^N\mathbf{w}_{ki}^H\mathbf{h}_{kj}s_{kj}
	+ \sqrt{\rho_{\text{ul}}} (\mathbf{w}_{ki}^H\mathbf{h}_{ki} - \mathbb{E}[\mathbf{w}_{ki}^H\mathbf{h}_{ki}])s_{ki}\notag  \\
	&+ \mathbf{w}_{ki}^H\mathbf{n}.
	\end{align}
	Then, the following ergodic \ac{UL} \ac{SE} for user $k$ is achievable
	\begin{equation}\label{eq:lower_bound}
	\textsf{SE}_{k} =  \frac{\tau_c-\tau_p}{\tau_c}\ \sum_{i=1}^N\log_2(1+\textsf{SINR}_{ki}) \quad [\text{bps/Hz}]
	\end{equation}
	where $\textsf{SINR}_{ki}$ is given by \eqref{eq:SINR} at the bottom of the page. 
		\begin{figure*}[b]
		\normalsize
		\hrulefill
		\vspace*{1pt}
		\begin{equation}\label{eq:SINR}
		\textsf{SINR}_{ki} =  \frac{\rho_{\text{ul}}\left | \mathbb{E}[\mathbf{w}_{ki}^H\mathbf{h}_{ki}]  \right |^2}{\sum_{m=1}^K\sum_{j=1}^N\rho_{\text{ul}}\mathbb{E}\left[\left | \mathbf{w}_{ki}^H\mathbf{h}_{mj} \right |^2\right] - \rho_{\text{ul}}\left | \mathbb{E}[\mathbf{w}_{ki}^H\mathbf{h}_{ki}] \right |^2 + \mathbb{E}[\|\mathbf{w}_{ki}\|^2]}.
		\end{equation}
		\normalsize
		\hrulefill
		\vspace*{1pt}
		\begin{align}\label{eq:xi}
		\xi_{ki} & = \text{tr}(\bm{\Phi}^k_{ii}) + \|\bar{\mathbf{h}}_{ki}\|^2\\[2pt]
		\zeta_{kimj} & = \text{tr}\left(\mathbf{\Phi}^k_{ii}\mathbf{C}^m_{jj}\right) + |\bar{\mathbf{h}}_{ki}^H\bar{\mathbf{h}}_{mj}|^2+
		\bar{\mathbf{h}}_{ki}^H\mathbf{R}^m_{jj}\bar{\mathbf{h}}_{ki} +\bar{\mathbf{h}}_{mj}^H\bm{\Phi}^k_{ii}\bar{\mathbf{h}}_{mj}  \label{eq:zeta} \\[2pt]
		&+	\text{tr}\left(\mathbf{\Phi}^m_{jj}\bm{\Phi}^k_{ii}\right), \quad \text{if} \ k\neq m \nonumber \\[2pt]
		&+\text{tr}\left(\mathbf{\Phi}^k_{ii}\bm{\Phi}^k_{ii}\right) + \left | \text{tr}(\bm{\Phi}^k_{ii})\right |^2  +2\text{tr}\left(\bm{\Phi}^k_{ii}\right)\|\bar{\mathbf{h}}_{ki}\|^2,  \quad \text{if} \ k= m \ \text{and} \ i=j\nonumber  \\[2pt]
		&+\sum_{l_1=1}^N\sum_{l_2=1}^N\left(\text{tr}(\tilde{\bm{\Phi}}^k_{l_1i}\tilde{\bm{\Phi}}^k_{jl_1})\text{tr}(\tilde{\bm{\Phi}}^k_{il_2}\tilde{\bm{\Phi}}^k_{l_2j})+\|\tilde{\bm{\Phi}}^k_{l_1i}\tilde{\bm{\Phi}}^k_{jl_2}\|_{\textsc{f}}^2\right)+2\text{Re}\left\{\text{tr}\left(\bm{\Phi}^k_{ij}\right)\bar{\mathbf{h}}_{ki}^H\bar{\mathbf{h}}_{kj}\right\}, \quad \text{if} \ k= m \ \text{and} \ i \neq j. \nonumber
		\end{align}
	\end{figure*}
Referring to \eqref{eq:post_signal}, only the part of the desired signal received over the average effective channel $\mathbb{E}[\mathbf{w}_{ki}^H\mathbf{h}_{ki}]$ is treated as the true desired signal in the detection. All interference terms has zero mean, and hence can be treated as uncorrelated noise in the detection. Since this represents the worst-case assumption when computing the mutual information, \eqref{eq:lower_bound} represents a lower bound on the ergodic sum capacity~\cite{mMIMO_book}. In the following theorem, we provide a closed-form expression for the SE in~\eqref{eq:lower_bound}.

	\begin{theorem}
		The SINR of stream $i$ of user $k$ under MMSE estimation and MR combining is given by
		\\
		\begin{equation}\label{eq:SINR_deterministic}
		\emph{\textsf{SINR}}_{ki} =  \frac{\rho_{\emph{ul}}|\xi_{ki}|^2}{\sum_{m=1}^K\sum_{j=1}^N\rho_{\emph{ul}}\zeta_{kimj} - \rho_{\emph{ul}}|\xi_{ki}|^2 + \xi_{ki}}
		\end{equation}
		\\
		where $\xi_{ki} \triangleq\mathbb{E}[\mathbf{w}_{ki}^H\mathbf{h}_{ki}] = \mathbb{E}[\|\mathbf{w}_{ki}\|^2]$ and $\zeta_{kimj} \triangleq\mathbb{E}[|\mathbf{w}_{ki}^H\mathbf{h}_{mj}|^2]$ are given by \eqref{eq:xi} and \eqref{eq:zeta} at the bottom of the page, respectively; $\tilde{\bm{\Phi}}^k_{ij}$ is the $(i,j)$-submatrix of $(\bm{\Phi}_{k})^{1/2}$, with $\bm{\Phi}_{k} \triangleq\mathbf{R}_k-\mathbf{C}_k$.
		\begin{proof}
			See Appendix.
		\end{proof}
	\end{theorem}
\subsection{Discussion}
		Our \ac{SE} analysis generalizes previous theoretical papers on massive \ac{MIMO} with correlated channels. More particularly, for correlated Rayleigh fading and single-antenna users, our result is identical with that in \cite[Ch.~4]{mMIMO_book}. Likewise, Theorem~1 complements the work in \cite{single_antenna_ricean1} by considering spatially correlated channels of multi-antenna users. Regarding the insights Theorem~1 can provide, an upper bound on~\eqref{eq:SINR_deterministic} can be constructed by utilizing the Rayleigh-Ritz theorem. Specifically, this bound is a function of the eigencoupling coefficients $[\bm{\Omega}_k]_{m,n}$ as well as the number of BS and user antennas $M$ and $N$, respectively, and can be used to show that the sum \ac{SE} will not grow without limit as $M$ increases when at least one coupling coefficient scales also as $\mathcal{O}(M)$; similar findings were reported in \cite{mMIMO_Ricean}. Providing a formal proof of the said bound is left for future work. However, this insight is consolidated in the subsequent numerical simulations. 
		
		\begin{figure*}
		\centering
		\begin{subfigure}{0.62\columnwidth}
			\includegraphics[width=\columnwidth]{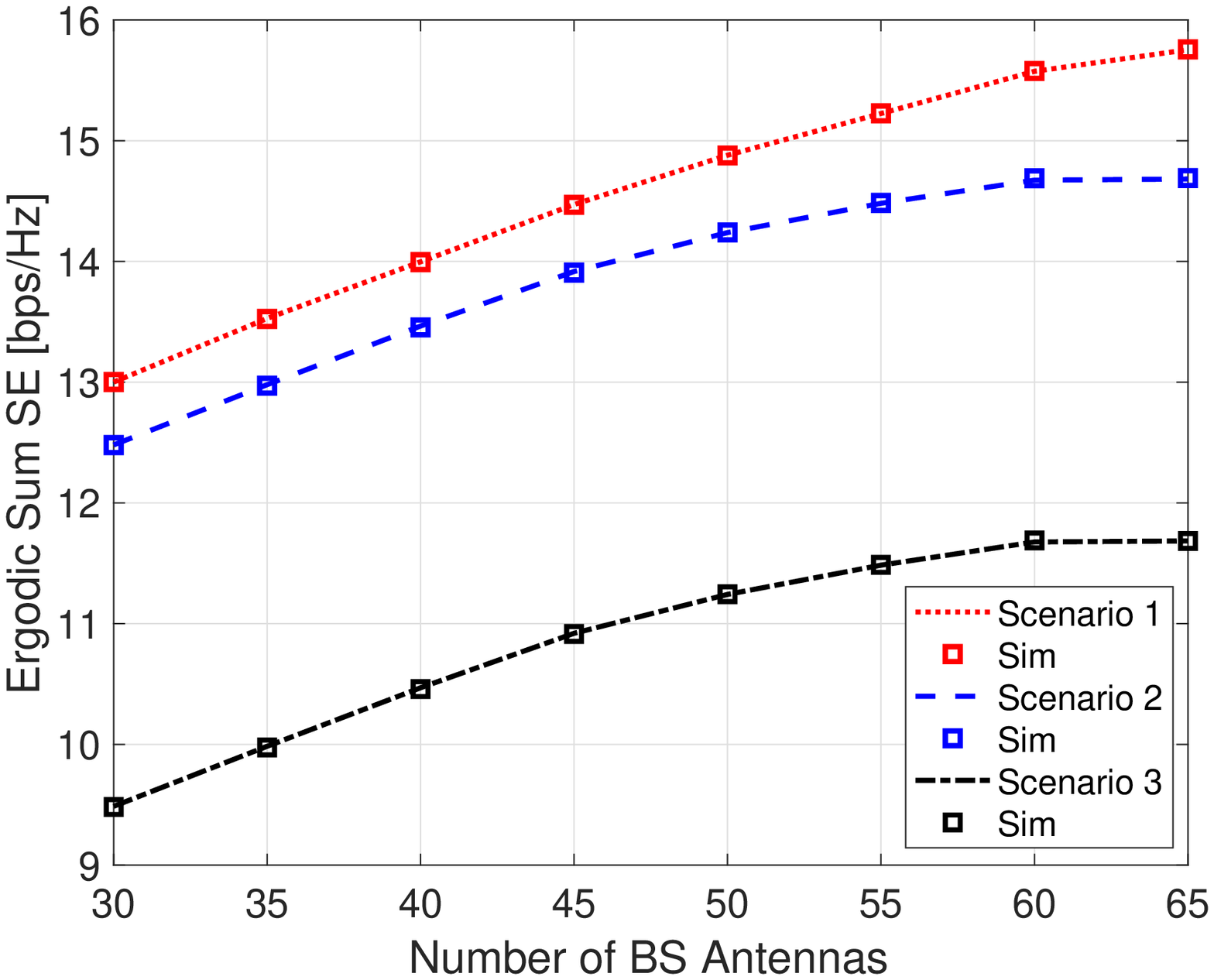}
			\caption{}
			\label{Fig:Result_1}
		\end{subfigure}\hfil
		\begin{subfigure}{0.62\columnwidth}
			\includegraphics[width=\columnwidth]{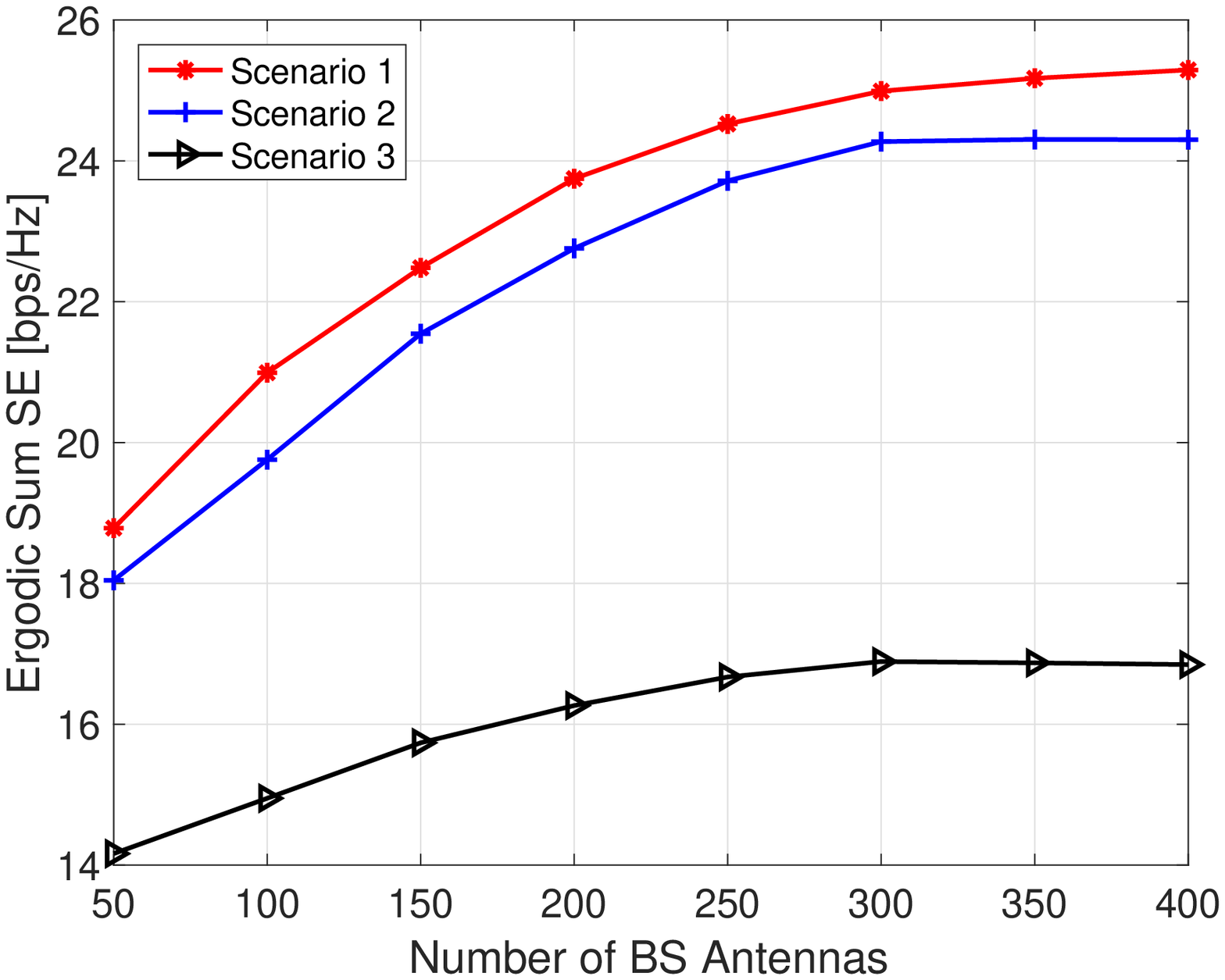}
			\caption{}
			\label{Fig:Result_2}
		\end{subfigure}
		\begin{subfigure}{0.62\columnwidth}
			\includegraphics[width=\columnwidth]{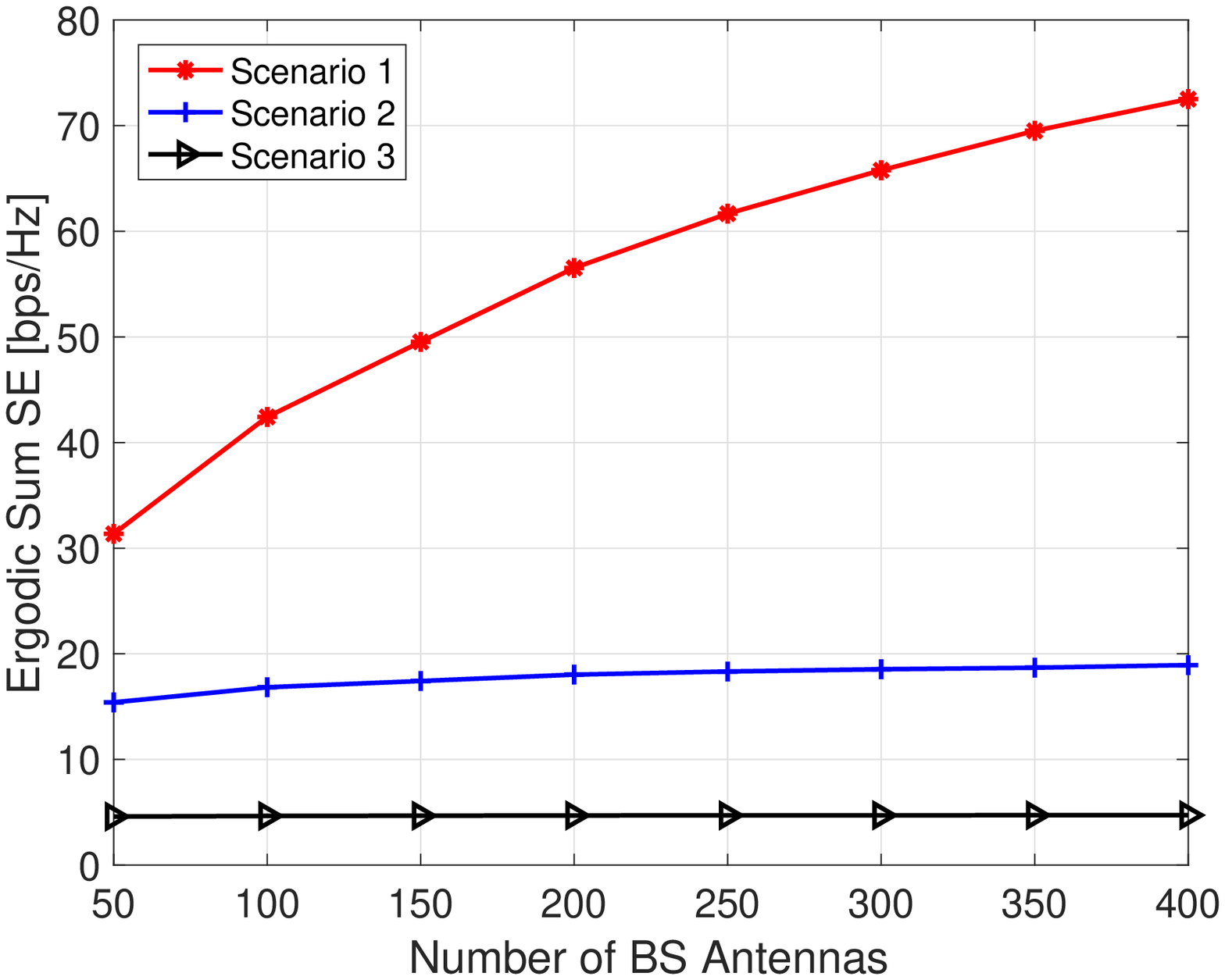}
			\caption{}
			\label{Fig:Result_3}
		\end{subfigure}
		\caption{Ergodic \ac{UL} sum SE for $8$ users and $2$ antennas/user: (a) validation of the closed-form expression, with square markers representing simulation data; (b) \ac{SE} under jointly correlated Ricean fading; (c) \ac{SE} under jointly correlated Rayleigh fading.}
		\label{Fig1}
	\end{figure*}
	
	\section{Numerical Results}
	In this section, we carry out Monte Carlo simulations to validate the closed-form \ac{SE} expression derived in Section \ref{section:se_analysis}. We further investigate the system performance under various propagation conditions. All simulation results are obtained for a $20$ MHz channel, a transmit \ac{SNR} of 10 dB, and a coherence block of $\tau_c = 500$ symbols. The Ricean factor $\kappa_k$ is drawn from $\mathcal{U}(2,4)$ [dB], which is a typical range of values in indoor deployments~\cite{3GPP_specs}. The \ac{LoS} channel component $\bar{\mathbf{H}}_k$  is expressed as \cite{los_model1}
	\begin{equation*}
	\bar{\mathbf{H}}_k = \sqrt{\frac{\kappa_k}{\kappa_k+1}}\  \mathbf{a}_{r}\left(\theta^r_k\right)\cdot\mathbf{a}_{t}\left(\theta^t_k\right)^T
	\end{equation*}
	where $\theta^t_k\sim\mathcal{U}(0,2\pi)$ is the angle-of-departure (AoD) of the \ac{LoS} path, $\theta^r_k\sim\mathcal{U}(0,2\pi)$ is the angle-of-arrival (AoA), and $\mathbf{a}_{t}\left(\theta^t_k\right)$ and $\mathbf{a}_{r}\left(\theta^r_k\right)$ are the transmit and receive array response vectors, respectively. We consider a uniform linear array (ULA) with half-wavelength antenna spacing at both \ac{BS} and user sides. The array response vectors are then given by 
	\begin{align*}
	\mathbf{a}_{t}\left(\theta^t_k\right) &\triangleq \left[1, e^{j\pi \sin\theta^t_k},\dots, e^{j\pi (N-1)\sin\theta^t_k}\right]^T,\\
	\mathbf{a}_{r}\left(\theta^r_k\right) &\triangleq \left[1, e^{j\pi \sin\theta^r_k},\dots, e^{j\pi (M-1)\sin\theta^r_k}\right]^T.
	\end{align*}
	\vspace{-0.5cm}
	\subsection{UL Spectral Efficiency}
	We assess the sum \ac{SE} of the system under three scenarios:
	\\[8pt]
	\textbf{Scenario 1}: $\bm{\Omega}_{k} = \frac{1}{\kappa_k+1}\mathbf{1}_{M\times N}$.
	\\[14pt]
	\textbf{Scenario 2}: $\bm{\Omega}_{k}=\begin{bmatrix} \frac{MN}{2(\kappa_k+1)} &  a\mathbf{1}_{1\times (N-1)} \\
	a & a\mathbf{1}_{(M-1)\times (N-1)}\end{bmatrix}$, where $a=\frac{MN}{2(\kappa_k+1)(MN-1)}$.
	\\[14pt]
	\textbf{Scenario 3}: $\bm{\Omega}_{k}= \frac{1}{\kappa_k+1}\bm{\lambda}_{k,r}\bm{\lambda}_{k,t}^T$, where  $\bm{\lambda}_{k,r}$ and $\bm{\lambda}_{k,t}$ are calculated for the coupling matrix of Scenario 2 using the constraints introduced in Section \ref{sec:system_model}.
	\\[8pt]
	The first scenario represents a rich scattering environment where the \ac{NLoS} channel components are treated as i.i.d. variables, i.e., uncorrelated Rayleigh fading. The second scenario represents a jointly correlated fading channel where there is an entry in each coupling matrix scaling as $\mathcal{O}(MN)$ \cite{WModel_SingleAntenna}. The last scenario is the Kronecker version of the correlated channel considered in Scenario 2. We stress that $\text{tr}(\mathbf{R}_k) = MN/(\kappa_k+1)$ holds in all the scenarios under investigation.

Figure \ref{Fig1}(\subref{Fig:Result_1}) shows the accuracy of the closed-form expression against simulation data for 10,000 channel realizations for jointly correlated Ricean fading. As we see, the Kronecker model yields lower \ac{SE} than the one described by the Weichselberger model, since it neglects the joint correlation feature of the channel, which agrees with the findings in \cite{wei_model}. Next, by capitalizing on the closed-form expression, we plot the sum \ac{SE} as a function of the number of \ac{BS} antennas under \ac{LoS} and \ac{NLoS} propagation. From Fig. \ref{Fig1}(\subref{Fig:Result_2}), we observe that correlation becomes irrelevant in the very large antenna regime due to \ac{LoS} propagation. Interestingly, due to the presence of \ac{LoS} components, the \ac{SE} gains from deploying a massive number of \ac{BS} antennas are not substantial. Likewise, under \ac{NLoS} propagation, Fig.~\ref{Fig1}(\subref{Fig:Result_3}) shows that the \ac{SE} does not grow without bound as the number of \ac{BS} antennas increases when at least one entry in the coupling matrix scales as $\mathcal{O}(MN)$. Similar results were reported in~\cite{WModel_SingleAntenna},~\cite{mMIMO_Ricean}.
 	\begin{figure}[H]
		\centering
		\includegraphics[width=0.65\columnwidth]{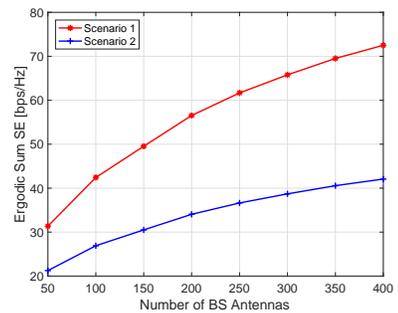}
		\caption{Ergodic \ac{UL} sum SE for $8$ users and $2$ antennas/user with user precoding and jointly correlated Rayleigh fading.}
		\label{Fig:Result_4}
	\end{figure}
	To circumvent this barrier, we examine the case where users employ precoding. Specifically, we consider that users precode their data with their transmit covariance matrix, i.e., transmit the signal $\sqrt{\rho_{\text{ul}}}\mathbf{U}_{k,t}^*\mathbf{s}_k$. By doing so, they decorrelate their data streams, and the full correlation matrix takes a block-diagonal form. Figure \ref{Fig:Result_4} depicts the sum \ac{SE} versus the number of \ac{BS} antennas for \ac{NLoS} and neglecting the Kronecker-type scenario. We observe that user precoding can enable the sum \ac{SE} to scale with the number of \ac{BS} antennas, even when the eigenmode coupling coefficients grow as $\mathcal{O}(MN)$. However, we point out that user precoding entails additional overhead since users have to learn their channel statistics. A similar approach was proposed in \cite{massive_mimo_multiantenna_users} for Kronecker-structured systems.
	
	\subsection{Multi-Antenna versus Single-Antenna Users}
	One fundamental question is when additional user antennas are beneficial. In the sequel, we investigate if it is better to serve a few multi-antenna users or many single-antenna users. We assess the sum \ac{SE} under the correlation Scenario 2 and a total number of data streams $KN=12$. We further consider the options $(12,1), (6,2)$, and $(4,3)$ for the pair $(K,N)$.
	\begin{figure}[H]
		\centering
		\includegraphics[width=0.65\columnwidth]{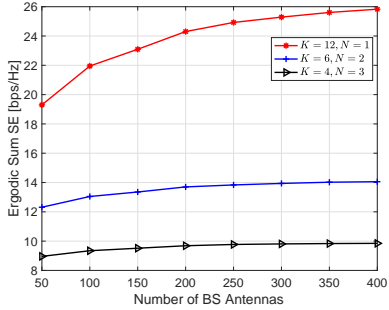}
		\caption{Ergodic \ac{UL} sum SE for different number of served users and user antennas.}
		\label{Fig:Result_5}
	\end{figure}
	From Fig. \ref{Fig:Result_5}, we notice that having single-antenna users is better in terms of the sum \ac{SE} compared to the multi-antenna case. We therefore conclude that serving users equipped with multiple antennas is not a good option if sum \ac{SE} is our objective. However, if we aim at improving other objectives such as per-user \ac{SE} or communication reliability, using multiple antennas at the users is a meaningful option.
	
	\section{Conclusions}
	We analyzed the performance of massive \ac{MIMO} with multi-antenna users under a generalized channel model. Specifically, we derived a closed-form expression for the \ac{UL} \ac{SE}, which is achievable using \ac{MR} combining and \ac{MMSE} estimation. Capitalizing on the closed-form expression, we pursued a rigorous study of the system performance. Our \ac{SE} analysis employs the Weichselberger model, and hence can incorporate the popular Kronecker and virtual channel representation models as special cases. Simulation results suggest that \ac{UL} \ac{SE} does not grow always without bound as the number of \ac{BS} antennas increases under jointly correlated fading, when at least one eigenmode coupling coefficient scales as $\mathcal{O}(MN)$. To circumvent this barrier, we proposed precoding at the users, which entails though additional overhead. In the case of omnidirectional \ac{UL} transmissions, the detrimental effect of the joint correlation structure of the channel can be alleviated by serving single-antenna users.
	
	\section*{Appendix: Proof of Theorem 1}
	The term $\mathbb{E}[\mathbf{w}_{ki}^H\mathbf{h}_{ki}]$ is written as 
	\begin{align}
	&\mathbb{E}[\mathbf{w}_{ki}^H\mathbf{h}_{ki}] = \mathbb{E}[\hat{\mathbf{h}}_{ki}^H\mathbf{h}_{ki}] =\mathbb{E}[\hat{\mathbf{h}}_{ki}^H\hat{\mathbf{h}}_{ki}] + \mathbb{E}[\hat{\mathbf{h}}_{ki}^H\hat{\mathbf{e}}_{ki}]\nonumber\\
	&\overset{(a)}{=} \mathbb{E}[\|\hat{\mathbf{h}}_{ki}\|^2] = \text{tr}(\mathbb{E}[\hat{\mathbf{h}}_{ki}\hat{\mathbf{h}}_{ki}^H])\nonumber\\
	&= \text{tr}(\mathbf{\Phi}^k_{ii} +\bar{\mathbf{h}}_{ki}\bar{\mathbf{h}}_{ki}^H)=\text{tr}(\mathbf{\Phi}^k_{ii}) + \|\bar{\mathbf{h}}_{ki}\|^2 \nonumber
	\end{align}
	where $(a)$ is because $\mathbb{E}[\hat{\mathbf{h}}_{ki}^H\hat{\mathbf{e}}_{ki}]=0$ under \ac{MMSE} estimation. We further have $\mathbb{E}[\|\mathbf{w}_{ki}\|^2] = \mathbb{E}[\|\hat{\mathbf{h}}_{ki}\|^2]=\mathbb{E}[\mathbf{w}_{ki}^H\mathbf{h}_{ki}]$, which completes the proof for $\xi_{ki}$.
	
	The second moment $\mathbb{E}[|\mathbf{w}_{ki}^H\mathbf{h}_{mj}|^2]$ is written as 
	\begin{align}\label{eq:eq0}
	\mathbb{E}[|\mathbf{w}_{ki}^H\mathbf{h}_{mj}|^2] &= \mathbb{E}[|\hat{\mathbf{h}}_{ki}^H\mathbf{h}_{mj}|^2] \nonumber\\
	&= \mathbb{E}[|\hat{\mathbf{h}}_{ki}^H(\hat{\mathbf{h}}_{mj} + \mathbf{e}_{mj})|^2]\nonumber\\
	& = \mathbb{E}[|\hat{\mathbf{h}}_{ki}^H\hat{\mathbf{h}}_{mj}|^2] + \mathbb{E}[|\hat{\mathbf{h}}_{ki}^H\mathbf{e}_{mj}|^2]
	\end{align}
	where \eqref{eq:eq0} follows the fact that the estimation error $\mathbf{e}_{mj}$ has zero mean, and is statistically independent of the channel estimates $\hat{\mathbf{h}}_{ki}$ and $\hat{\mathbf{h}}_{mj}$. Based on this property, we have that 
	\begin{align}\label{eq:eq1}
	\mathbb{E}[|\hat{\mathbf{h}}_{ki}^H\mathbf{e}_{mj}|^2] &= \mathbb{E}\left[\text{tr}\left(\hat{\mathbf{h}}_{ki}^H\mathbf{e}_{mj}\mathbf{e}_{mj}^H\hat{\mathbf{h}}_{ki}\right)\right]\nonumber\\
	&=\text{tr}\left(\mathbb{E}\left[\hat{\mathbf{h}}_{ki}\hat{\mathbf{h}}_{ki}^H\mathbf{e}_{mj}\mathbf{e}_{mj}^H\right]\right)\nonumber\\
	&=\text{tr}\left(\mathbb{E}\left[\hat{\mathbf{h}}_{ki}\hat{\mathbf{h}}_{ki}^H\right]\mathbb{E}\left[\mathbf{e}_{mj}\mathbf{e}_{mj}^H\right]\right)\nonumber\\
	&=\text{tr}(\mathbf{\Phi}^k_{ii}\mathbf{C}^m_{jj}) + \bar{\mathbf{h}}_{ki}^H\mathbf{C}^m_{jj}\bar{\mathbf{h}}_{ki}.
	\end{align}
	Combining \eqref{eq:eq0} and \eqref{eq:eq1} yields
	\begin{equation}\label{eq:eq2}
	\mathbb{E}[|\mathbf{w}_{ki}^H\mathbf{h}_{mj}|^2] =  \mathbb{E}[|\hat{\mathbf{h}}_{ki}^H\hat{\mathbf{h}}_{mj}|^2] + \text{tr}(\mathbf{\Phi}^k_{ii}\mathbf{C}^m_{jj}) + \bar{\mathbf{h}}_{ki}^H\mathbf{C}^m_{jj}\bar{\mathbf{h}}_{ki}.
	\end{equation}
	In \eqref{eq:eq2}, the term $\mathbb{E}[|\hat{\mathbf{h}}_{ki}^H\hat{\mathbf{h}}_{mj}|^2]$ is evaluated explicitly for the following three cases:
	\\[5pt]
	Case $k\neq m$. The vectors $\hat{\mathbf{h}}_{ki}$ and $\hat{\mathbf{h}}_{mj}$ are statistically independent (i.e., channels of different users). Hence, algebraic manipulations akin to \eqref{eq:eq1} yield
	\begin{equation}\label{eq:eq3}
	\begin{multlined}
	\mathbb{E}[|\hat{\mathbf{h}}_{ki}^H\hat{\mathbf{h}}_{mj}|^2] = \text{tr}(\mathbf{\Phi}^k_{ii}\mathbf{\Phi}^m_{jj}) + \bar{\mathbf{h}}_{ki}^H\mathbf{\Phi}^m_{jj}\bar{\mathbf{h}}_{ki} \\+ \bar{\mathbf{h}}_{mj}^H\mathbf{\Phi}^k_{ii}\bar{\mathbf{h}}_{mj} + |\bar{\mathbf{h}}_{ki}^H\bar{\mathbf{h}}_{mj}|^2.
	\end{multlined}
	\end{equation}
	Combining \eqref{eq:eq2}, \eqref{eq:eq3}, and the identity $\mathbf{\Phi}^m_{jj} = \mathbf{R}^m_{jj}-\mathbf{C}^m_{jj}$, gives the desired result.
	\\[5pt]
	Case $ k=m$ and $i=j$. For ease of notation, we drop the subscript $k$. This second moment is computed by writing $\hat{\mathbf{h}}_{i}$ as $\hat{\mathbf{h}}_{i} = \bm{\Phi}_{ii}^{1/2}\mathbf{x} + \bar{\mathbf{h}}_{i}$, where $\mathbf{x}\sim\mathcal{CN}(\mathbf{0},\mathbf{I})$. We then have
	\begin{align}\label{eq:eq5}
	\mathbb{E}&[|\hat{\mathbf{h}}_{i}^H\hat{\mathbf{h}}_{i}|^2] = \mathbb{E}\left[|(\bm{\Phi}_{ii}^{1/2}\mathbf{x} + \bar{\mathbf{h}}_{i})^H(\bm{\Phi}_{ii}^{1/2}\mathbf{x} + \bar{\mathbf{h}}_{i})|^2\right]\nonumber\\[1pt]
	&=\mathbb{E}\left[| \underbrace{\mathbf{x}^H\bm{\Phi}_{ii}\mathbf{x}}_{a} + \underbrace{\mathbf{x}^H\bm{\Phi}_{ii}^{1/2}\bar{\mathbf{h}}_{i}}_{b} + \underbrace{\bar{\mathbf{h}}_{i}^H\bm{\Phi}_{ii}^{1/2}\mathbf{x}}_{c} + \underbrace{\|\bar{\mathbf{h}}_{i}\|^2}_{d} |^2\right].
	\end{align}
	The individual terms are determined as follows. $\mathbb{E}[aa^*] = \mathbb{E}[|\mathbf{x}^H\mathbf{\Phi}_{ii}\mathbf{x}|^2] = |\text{tr}(\mathbf{\Phi}_{ii})|^2 + \text{tr}(\mathbf{\Phi}_{ii}^2)$, which is a standard matrix identity \cite{mMIMO_book}; $\mathbb{E}[ab^*] = \mathbb{E}[ba^*]= 0$ because it involves odd order of moments (i.e., multivariate Gaussian distribution); likewise $\mathbb{E}[ac^*] = \mathbb{E}[ca^*]= 0$; $\mathbb{E}[ad^*]= \mathbb{E}[da^*] = \text{tr}(\mathbf{\Phi}_{ii})\|\bar{\mathbf{h}}_i\|$; $\mathbb{E}[bb^*]=\bar{\mathbf{h}}_i^H\mathbf{\Phi}_{ii}\bar{\mathbf{h}}_i$; $\mathbb{E}[bc^*]=\mathbb{E}[cb^*] = 0$ due to the circular symmetry of $\mathbf{x}$, i.e., $\mathbb{E}[\mathbf{x}\mathbf{x}^T]=0$; $\mathbb{E}[bd^*]=\mathbb{E}[db^*] =\mathbb{E}[cd^*] = \mathbb{E}[dc^*] = 0$ because $\mathbf{x}$ has zero mean; $\mathbb{E}[cc^*]=\bar{\mathbf{h}}_i^H\mathbf{\Phi}_{ii}\bar{\mathbf{h}}_i$; and $\mathbb{E}[dd^*] = \|\bar{\mathbf{h}}_i\|^4$. Thus,
	\begin{align}\label{eq:eq4}
	\mathbb{E}[|\hat{\mathbf{h}}_{i}^H\hat{\mathbf{h}}_{i}|^2] &= |\text{tr}(\mathbf{\Phi}_{ii})|^2 + \text{tr}(\mathbf{\Phi}_{ii}^2) + 2\bar{\mathbf{h}}_i^H\mathbf{\Phi}_{ii}\bar{\mathbf{h}}_i \nonumber\\
	&+ 2\text{tr}(\mathbf{\Phi}_{ii})\|\bar{\mathbf{h}}_i\| + \|\bar{\mathbf{h}}_i\|^4.
	\end{align}
	Combining \eqref{eq:eq2} and \eqref{eq:eq4} completes the proof. 
	\\[5pt]
	Case $ k=m$ and $i\neq j$. For ease of notation, we drop the subscript $k$ hereafter. In this case, we cannot decompose the vectors $\hat{\mathbf{h}}_{i}$ and $\hat{\mathbf{h}}_{j}$ in terms of a single complex normal Gaussian vector, because the circular symmetry property will not be preserved. According to \eqref{eq:correlation_blockform} and \eqref{eq:covariance_error_blockform}, we can express $\hat{\mathbf{h}}_i$ and $\hat{\mathbf{h}}_j$ into the equivalent form $\hat{\mathbf{h}}_i = \sum_{l=1}^N\tilde{\mathbf{\Phi}}_{il}\mathbf{x}_l + \bar{\mathbf{h}}_i$ and $\hat{\mathbf{h}}_j = \sum_{l=1}^N\tilde{\mathbf{\Phi}}_{jl}\mathbf{x}_l + \bar{\mathbf{h}}_j$, respectively, where $\mathbf{x}_l\sim\mathcal{CN}(\mathbf{0},\mathbf{I})$, and $\tilde{\mathbf{\Phi}}_{il}$ denotes the $(i,l$)-th submatrix of $\mathbf{\Phi}^{1/2}$; to see this write $\text{vec}(\hat{\mathbf{H}}) = \mathbf{\Phi}^{1/2}[\mathbf{x}_1^T,\dots,\mathbf{x}_N^T]^T + \text{vec}(\bar{\mathbf{H}}) $. Also, the following two identities hold
	\begin{align}\label{identity1}
	\mathbf{\Phi}_{ii} &= \sum_{l=1}^N\tilde{\mathbf{\Phi}}_{il}\tilde{\mathbf{\Phi}}_{il}^H = \sum_{l=1}^N\tilde{\mathbf{\Phi}}_{il}\tilde{\mathbf{\Phi}}_{li},\\ \label{identity2}
	\mathbf{\Phi}_{ij} &= \sum_{l=1}^N\tilde{\mathbf{\Phi}}_{il}\tilde{\mathbf{\Phi}}_{jl}^H = \sum_{l=1}^N\tilde{\mathbf{\Phi}}_{il}\tilde{\mathbf{\Phi}}_{lj}.
	\end{align}
	We now have
	\begin{align}
	\mathbb{E}&[|\hat{\mathbf{h}}_{i}^H\hat{\mathbf{h}}_{j}|^2] =\nonumber\\
	& \mathbb{E}\left[ \left | \left(\sum_{l_1=1}^N\tilde{\bm{\Phi}}_{il_1}\mathbf{x}_{l_1} + \bar{\mathbf{h}}_{i}\right)^H
	\left(\sum_{l_2=1}^N\tilde{\bm{\Phi}}_{jl_2}\mathbf{x}_{l_2} + \bar{\mathbf{h}}_{j}\right)\right |^2 \right] \nonumber.
	\end{align}
	Expanding the product yields an expression akin to \eqref{eq:eq5}. Utilizing the identities in \eqref{identity1} and \eqref{identity2} and doing some algebra yields several terms similar to the case $i=j$, with the exception of $\mathbb{E}[ad^*]=  \text{tr}(\mathbf{\Phi}_{ij})\bar{\mathbf{h}}_i^H\bar{\mathbf{h}}_j$, $\mathbb{E}[da^*] = (\mathbb{E}[ad^*])^*$, and $\mathbb{E}[dd^*] = |\bar{\mathbf{h}}^H_i\bar{\mathbf{h}}_j|^2$. The only term remaining unknown is
	\begin{align*}
	&\mathbb{E}[aa^*] =\nonumber\\
	& \mathbb{E}\left[\left( \sum_{(l_1,l_2)}\mathbf{x}_{l_1}^H\tilde{\bm{\Phi}}_{l_1i}\tilde{\bm{\Phi}}_{jl_2}\mathbf{x}_{l_2} \right)
	\left(\sum_{(l_3,l_4)}\mathbf{x}_{l_3}^H\tilde{\bm{\Phi}}_{l_3i}\tilde{\bm{\Phi}}_{jl_4}\mathbf{x}_{l_4}  \right)^*\right].
	\end{align*}
	Due to lack of space, we only sketch the proof of that term. If $l_1=l_2=l_3=l_4$, we get the standard identity from \cite{mMIMO_book}, that is $\sum_{l}\mathbb{E}[|\mathbf{x}_l^H\tilde{\mathbf{\Phi}}_{li}\tilde{\mathbf{\Phi}}_{jl}\mathbf{x}_l|^2] = \sum_{l} |\text{tr}(\tilde{\mathbf{\Phi}}_{li}\tilde{\mathbf{\Phi}}_{jl})|^2 + \|\tilde{\mathbf{\Phi}}_{li}\tilde{\mathbf{\Phi}}_{jl}\|^2_\textsc{f}$. If at least one of the $l_i, i=1,\dots,4$, is different from the others, then we have $\mathbb{E}[aa^*]=0$ due to the circular symmetry property and the mean value of the individual Guassian vectors. If $l_1=l_2$ and $l_3=l_4$, $\mathbb{E}[aa^*]=\sum_{l_1,l_3}\text{tr}(\tilde{\mathbf{\Phi}}_{l_1i}\tilde{\mathbf{\Phi}}_{jl_1})\text{tr}(\tilde{\mathbf{\Phi}}_{il_3}\tilde{\mathbf{\Phi}}_{l_3j})$. If $l_1=l_3$ and $l_2=l_4$, $\mathbb{E}[aa^*]=\sum_{l_1,l_2}\|\mathbf{\tilde{\Phi}}_{l_1i}\tilde{\mathbf{\Phi}}_{jl_3}\|^2_\textsc{f}$. Finally, if $l_1=l_4$ and $l_2=l_3$, $\mathbb{E}[aa^*]=0$ due to the circular symmetry property. Putting all the individual terms together completes the proof.
	
	\section*{Acknowledgments}
	The work done by K. Dovelos and B. Bellalta was supported by grants PGC2018-099959-B-I00 (MCIU/AEI/FEDER,UE), and 2017-SGR-1188. The work of M. Matthaiou was supported by EPSRC, UK, under grant EP/P000673/1 and by the  RAEng/The Leverhulme Trust Senior Research Fellowship LTSRF1718$\backslash$14$\backslash$2. The work of H. Q. Ngo was supported by the UK Research and Innovation Future Leaders Fellowships under Grant MR/S017666/1.

\end{document}